\documentstyle[12pt,epsf]{article}

\newcommand{\nc}{\newcommand}
\nc{\fdiag}{0}
\nc{\bg}{B. Grz${{{\rm a}_{}}_{}}_{\hskip -0.18cm\varsigma}$dkowski}
\nc{\lsp}{\;\;\;\;\;\;\;\;}
\nc{\beq}{\begin{equation}}   \nc{\eeq}{\end{equation}}
\nc{\bea}{\begin{eqnarray}}   \nc{\eea}{\end{eqnarray}}
\nc{\baa}{\begin{array}}      \nc{\eaa}{\end{array}}
\nc{\bit}{\begin{itemize}}    \nc{\eit}{\end{itemize}}
\nc{\ben}{\begin{enumerate}}  \nc{\een}{\end{enumerate}}
\nc{\bce}{\begin{center}}     \nc{\ece}{\end{center}}
\nc{\non}{\nonumber}
\nc{\lumun}{\;{\hbox {pb}^{-1}}{\hbox {yr}^{-1}}}
\nc{\hc}{\hbox {h.c.}}
\nc{\re}{\hbox {Re}}
\nc{\im}{\hbox {Im}}
\nc{\etal}{\hbox{et al.}}
\nc{\ra} {\rightarrow}
\nc{\cw}{\cos\theta_W}        \nc{\sw}{\sin\theta_W}
\nc{\ttbar}{t\bar{t}}
\nc{\bbbar}{b\bar{b}}
\nc{\tanb} {\tan \beta}
\nc{\twbdec} {t\rightarrow W^+ b}
\nc{\tbwbdec} {\bar{t} \rightarrow W^- \bar{b}}
\nc{\hprod} {e^+e^- \ra Z^\ast \ra H Z}
\nc{\epem} {e^+e^-}
\nc{\wpwm} {W^+W^-}
\nc{\tbar} {\bar{t}}
\nc{\bbar} {\bar{b}}
\nc{\wpp} {W^+}
\nc{\mt}{m_t}
\nc{\mts}{m_t^2}
\nc{\mw} {m_W}
\nc{\mws} {m_W^2}
\nc{\mz} {m_Z}
\nc{\mzs} {m_Z^2}
\nc{\mh} {m_H}
\nc{\mhs} {m_H^2}
\nc{\ma} {m_A}
\nc{\mas} {m_A^2}
\nc{\hdec}{H \ra t\bar{t}}
\nc{\ttbardec}{\ttbar \ra W^+W^-\bbbar}
\nc{\po}{\Phi_1}
\nc{\pod}{\Phi_1^\dagger}
\nc{\pht}{\Phi_2}
\nc{\phtd}{\Phi_2^\dagger}
\nc{\phtt}{{\tilde{\Phi}}_2}
\nc{\popo}{\po^\dagger\po}
\nc{\phtpt}{\pht^\dagger\pht}
\nc{\popt}{\po^\dagger\pht}
\nc{\phtpo}{\pht^\dagger\po}
\nc{\sq}{\sqrt{2}}
\nc{\nsd} {N_{SD}}
\nc{\ntt} {N_{tt}}

\def\lsim{\mathrel{\raise.3ex\hbox{$<$\kern-.75em\lower1ex\hbox{$\sim$}}}}
\def\gsim{\mathrel{\raise.3ex\hbox{$>$\kern-.75em\lower1ex\hbox{$\sim$}}}}

\def\pbi{~{\rm pb}^{-1}}

\def\gev{\,{\rm GeV}}

\def\wt{\widetilde}

\def\thdm{2HDM}
\def\tds{2D1S}

\def\br{BR}

\def\rts{\sqrt s}
\def\eps{\epsilon}
\def\h{h}
\def\a{a}
\def\mh{m_{\h}}
\def\ma{m_{\a}}

\def\etal{{\it et al.}}
\def\epem{e^+e^-}

\def\lsim{\mathrel{\raise.3ex\hbox{$<$\kern-.75em\lower1ex\hbox{$\sim$}}}}
\def\gsim{\mathrel{\raise.3ex\hbox{$>$\kern-.75em\lower1ex\hbox{$\sim$}}}}
\def\@versim#1#2{\vcenter{\offinterlineskip
        \ialign{$\m@th#1\hfil##\hfil$\crcr#2\crcr\sim\crcr } }}
\def\nsd{N_{SD}}

\def\pbi{~{\rm pb}^{-1}}

\def\gev{\,{\rm GeV}}

\def\wt{\widetilde}

\def\sq{\wt q}

\def\hsm{h_{\rm SM}}
\def\mhsm{m_{\hsm}}
\def\hl{h^0}
\def\hh{H^0}
\def\ha{A^0}

\def\hpm{H^{\pm}}
\def\mhl{m_{\hl}}
\def\mhh{m_{\hh}}
\def\mha{m_{\ha}}

\def\tanb{\tan\beta}

\def\mt{m_t}

\def\mz{m_Z}
\def\mw{m_W}

\def\wp{W^+}
\def\wm{W^-}

\def\h{h}
\def\mh{m_{\h}}

\begin{document}
%
\font\fortssbx=cmssbx10 scaled \magstep2
\medskip
$\vcenter{
\hbox{\fortssbx University of California - Davis}
}$
\hfill
$\vcenter{
\hbox{\bf UCD-97-11} 
\hbox{\bf SCIPP/97-09}
\hbox{\bf IFT-30-96}
\hbox{April, 1997}
}$
\vspace*{1cm}
\begin{center}
{\large{\bf LEP Limits on CP-Violating Non-Minimal Higgs Sectors}}\\
\rm
\vspace*{1cm}
{\bf  J.F. Gunion$^a$,
B. Grzadkowski$^b$, H.E. Haber$^c$ and J. Kalinowski$^{b,d}$\\
\vspace*{1cm}
{$^a$ \em Davis Institute for High Energy Physics, 
University of California, Davis, CA, USA }\\
{$^b$ \em Institute of Theoretical Physics, Warsaw University, 
Warsaw, Poland}\\
{$^c$ \em  University of California, Santa Cruz, CA, USA}\\
{$^d$ \em Deutsches Elektronen-Synchrotron, DESY, Hamburg, Germany}}
\end{center}
\begin{abstract}
We derive a sum rule which shows how to extend LEP limits
on the masses of the lightest CP-even and CP-odd Higgs
bosons of a CP-conserving two-Higgs-doublet model to any
two Higgs bosons of a general CP-violating two-Higgs-doublet model.
We generalize the analysis to a Higgs sector consisting
of an arbitrary number of Higgs doublets and singlets,
giving explicit limits for the CP-conserving and CP-violating
two-doublet plus one-singlet Higgs sectors.
\end{abstract}
\vspace{.5in}

\begin{figure}[ht]
\leavevmode
\epsfxsize=4.50in
\centerline{\epsffile{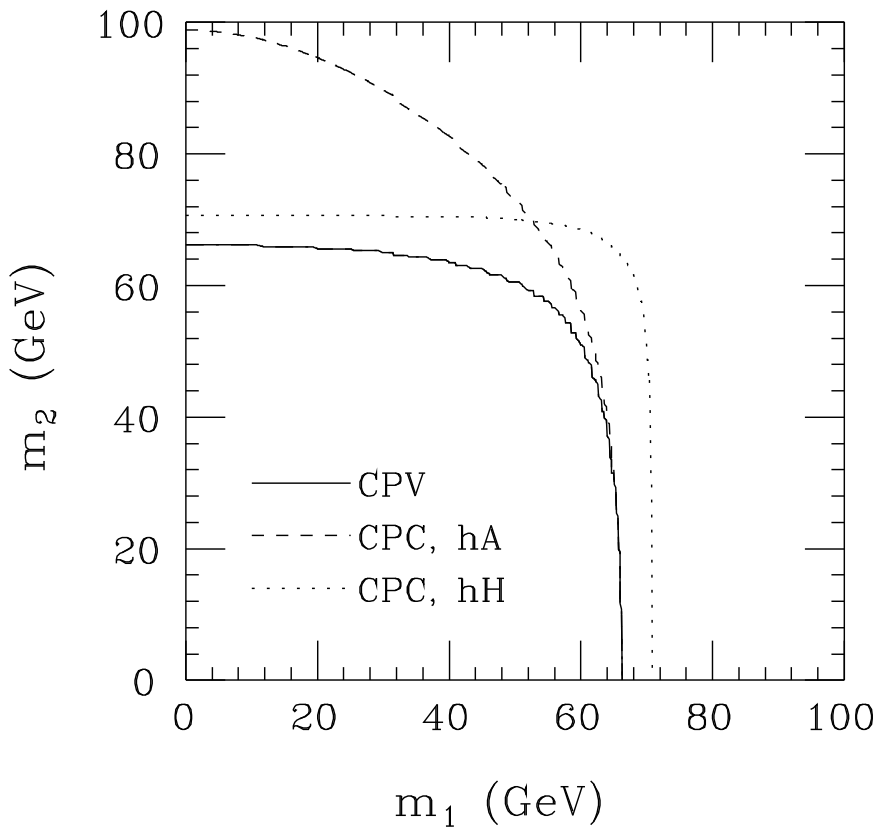}}
\bigskip
\caption{The LEP 95\% confidence level boundaries (based on
our use of the latest ALEPH results \protect\cite{alephnew}
and earlier LEP1 results as contained in Refs.~\protect\cite{l3new,sopczak})
in the $(m_1,m_2)$
plane for three two-Higgs-doublet model cases: (i) the CP-conserving (CPC)
model where $\h_1=\hl$, $\h_2=\hh$
(dotted curve); (ii) the CPC
model where $\h_1=\hl$, $\h_2=\ha$ (dashed curve); (iii) 
the CP-violating (CPV) model
where $\h_1$ and $\h_2$ are the two lightest neutral Higgs bosons 
(solid curve).
This figure is based on including just two constraints:
a particular choice of masses is excluded (a) if some $ZZ\h_i$ coupling must lie
above the 95\% CL upper limit for the assumed $m_{\h_i}$ or (b) if
one or more events are expected in $\epem\to\h_i\h_j$ production
at the 95\% CL for some choice of $i\neq j$. For details, see 
the Appendix.}
\label{fig95cl}
\end{figure}

Models of electroweak symmetry breaking driven by elementary scalar 
dynamics predict the existence of one or more physical Higgs bosons.
One can use LEP data to place significant bounds on Higgs boson masses.
The minimal model
consists of a one-doublet Higgs sector as employed in the 
Standard Model (SM), which
gives rise to a single CP-even scalar Higgs, $\hsm$. The absence of any 
$\epem\to Z \hsm$ signal in LEP1 data (where the $Z$ is virtual) and LEP2 data
(where the $Z$ is real) 
translates into a lower limit on $\mhsm$ which has been increasing
as higher energy data becomes available. For example, the latest
ALEPH data implies $\mhsm\gsim 70.7\gev$~\cite{alephnew}. 
Ultimately, the strongest limit will be obtained by combining the ALEPH
data with similar results from the L3, OPAL and DELPHI experiments.
The simplest and most attractive generalization of the SM Higgs sector is a
two-Higgs-doublet model (\thdm), the CP-conserving version of which has
received considerable attention, especially in the context of the minimal
supersymmetric model (MSSM) \cite{hhg}. 
A CP-conserving two-Higgs-doublet model predicts the
existence of two neutral CP-even Higgs bosons ($\hl$ and $\hh$, with
$\mhl\leq\mhh$), one neutral CP-odd Higgs ($\ha$) and a charged Higgs pair
($\hpm$). The negative results of Higgs boson searches at LEP can be
formulated  as restrictions on the parameter space of this and more general
Higgs sector models. 
For the most general CP-conserving two-doublet model one can
exclude the $(\mhl,\mha)$ and $(\mhl,\mhh)$ regions shown
in Fig.~\ref{fig95cl} on the basis of $\epem\to Z\hl$, 
$\epem\to Z\hh$ and $\epem\to \hl\ha$ 
event rate limits,\footnote{Stronger limits
are possible in models such as the MSSM where there
are relations between the Higgs masses and their couplings.} as explained
in the Appendix. The
ability to exclude the illustrated $(\mhl,\mha)$ region
derives from a coupling constant sum rule
which implies that the two production 
processes, $\epem\to Z \hl$ and $\epem\to \hl\ha$, cannot 
both be simultaneously
suppressed when kinematically allowed. Similarly, the illustrated
$(\mhl,\mhh)$ region
is excluded by virtue of a second sum rule which 
implies that the couplings responsible for the
$\epem\to Z\hl$ and $\epem\to Z \hh$ processes cannot be simultaneously
suppressed.

However, there is no reason
to assume that the Higgs sector is CP-conserving; CP-violation is still very
much a mystery from both an experimental and theoretical point of view,
and need not be entirely a consequence of the complex Yukawa couplings built
into the Kobayashi-Maskawa matrix \cite{km}. The possibility
that CP-violation derives largely from the Higgs sector
is especially intriguing \cite{weinberg}. In a general \thdm,
CP-violation can arise either explicitly or spontaneously and leads to
three electrically neutral physical Higgs mass eigenstates, $\h_i$
($i=1,2,3$) that have undefined CP properties. To date,
only the obvious limit on each $ZZ\h_i$ coupling as a function of $m_{\h_i}$,
deriving from non-observation of $\epem\to Z\h_i$ at LEP, has been noted.
In this Letter, we show that the
coupling constant sum rules appropriate in the CP-conserving model
can be generalized in the CP-violating case to yield a single sum rule
that requires at least one of the $ZZ\h_i$,
$ZZ\h_j$ and $Z \h_i\h_j$ ({\it any} $i\neq j$, $i,j=1,2,3$)
couplings to be substantial in size.
The LEP 95\% confidence level exclusion region in the 
$(m_{\h_i},m_{\h_j})$ plane that results from the general
sum rule is quite significant, as 
illustrated in Fig.~\ref{fig95cl} using $i=1,j=2$ notation.

In this Letter, we also present an analysis of both the CP-conserving
and the CP-violating versions of the two-doublet plus one-singlet (\tds)
extension of the \thdm. 
A \tds\ model yields five neutral Higgs bosons.
We derive sum rules that can be used to demonstrate
that LEP data excludes the possibility that three of these
neutral Higgs bosons can be light.  (There is no sum rule
which allows exclusion of the possibility that only
two of the neutral Higgs bosons of the \tds\ model are light.)
The 95\% confidence level boundaries in three-Higgs-boson mass space
for the CP-conserving and CP-violating cases, based on
the procedures described in the Appendix,
are presented in Figs.~\ref{fig2d1scpc}
and \ref{fig2d1scpv}, respectively.

We now turn to a derivation of the sum rules required and a discussion
of how they lead to the experimental constraints outlined above.
In the \thdm, the two
complex neutral Higgs fields contain four neutral degrees of freedom.
One is eaten by the $Z$
gauge boson; the others mix to yield three physical neutral Higgs bosons,
$\h_i$ ($i=1,2,3$). We shall denote their couplings to the $Z$ boson by
\begin{equation}
g_{ZZ\h_i}\equiv {g\mz\over c_W}C_i\,,\quad g_{Z\h_i\h_j}\equiv {g\over 2
c_W}C_{ij}\,,\quad g_{ZZ\h_i\h_j}={g^2\over 2c_W^2}\delta_{ij}\,,
\label{zcoups}
\end{equation}
where $C_i$ and $C_{ij}=C_{ji}$ ($i\neq j$)~\footnote{For $i=j$, $C_{ij}=0$
by Bose symmetry.} are model-dependent coupling strengths
and $c_W\equiv \cos\theta_W$.

In the CP-conserving \thdm, the $\hl$ and $\hh$ are
mixtures of the real parts of the neutral Higgs fields
(the diagonalizing mixing angle is denoted by $\alpha$) 
while the CP-odd state, $\ha$,
derives from the imaginary components not eaten by the $Z$.
One finds $C_{\hl}=C_{\hh\ha}=\sin(\beta-\alpha)$, 
$C_{\hh}=C_{\hl\ha}=\cos(\beta-\alpha)$ and $C_{\ha}=C_{\hl\hh}=0$; 
$\tanb=v_2/v_1$ is the ratio of the
vacuum expectation values for the neutral components of the two
Higgs doublet fields. For any two Higgs bosons, we wish to obtain 
an excluded mass region that does not depend upon 
knowledge of the third Higgs boson.
There are two pairs of interest: (a) $\hl,\hh$ and (b) $\hl,\ha$.
In case (a) the excluded region in $(\mhl,\mhh)$
parameter space is illustrated by the dotted curve in Fig.~\ref{fig95cl}.
Inside the excluded region,\footnote{We have plotted
this region symmetrically, even though by definition we
should only consider $\mhh\geq\mhl$.}
non-observation of $\epem\to Z \hl, Z \hh$ at LEP implies 
95\% CL limits on $C_{\hl}^2$ and $C_{\hh}^2$ such
that $C_{\hl}^2+C_{\hh}^2<1$, whereas the relevant couplings obey the sum rule
\beq
C_{\hl}^2+C_{\hh}^2=\sin^2(\beta-\alpha)+\cos^2(\beta-\alpha)=1\,.
\label{hHsr}
\eeq
In case (b) the excluded region in $(\mhl,\mha)$ parameter space
is indicated by the dashed line in Fig.~\ref{fig95cl}.
Inside the excluded region, failure to observe $\epem\to Z \hl$ and
$\epem\to \hl\ha$ events implies 95\% CL
upper limits on $C_{\hl}^2$ and $C_{\hl\ha}^2$, respectively, such that 
$C_{\hl}^2+C_{\hl\ha}^2<1$, whereas these couplings obey 
the sum rule
\beq
C_{\hl}^2+C_{\hl\ha}^2=\sin^2(\beta-\alpha) +\cos^2(\beta-\alpha)=1\,.
\label{hAsr}
\eeq       
Thus, LEP data exclude the possibility that both the scalar and
pseudoscalar Higgs bosons of a CP-conserving \thdm\ are light. 
The asymmetry of the $(\mhl,\mha)$ excluded region arises as follows.
If $\mhl$ is small, then non-observation of $ Z \hl$
events implies that $C_{\hl}^2$ is limited
to very small values, implying via Eq.~(\ref{hAsr}) that $C_{\hl\ha}^2\sim 1$.
Large $\mha$ is then required for the predicted
number of $\hl\ha$ events to be consistent with 0.
In contrast, if $\mhl$ is close to (or above) $\rts-\mz$, then
$ Z \hl$ events would not be observed
even if $C_{\hl}^2=1$: at the same time, $C_{\hl}^2=1$
implies via Eq.~(\ref{hAsr}) that
$C_{\hl\ha}^2=0$ so that there is no constraint
from non-observation of $\hl\ha$ events.

Below, we shall demonstrate that if CP is not conserved,
the sum rules of Eqs.~(\ref{hHsr}) and (\ref{hAsr}) can
be generalized\footnote{Note that Eq.~(\ref{newsr}) reduces to Eq.~(\ref{hAsr}) 
in the CP-conserving limit
when we identify $\h_i=\hl$ and $\h_j=\ha$ and use $C_{\ha}=0$.
Similarly, Eq.~(\ref{newsr}) reduces to Eq.~(\ref{hHsr}) 
in the CP-conserving limit
when we identify $\h_i=\hl$ and $\h_j=\hh$ and use $C_{\hl\hh}=0$.}
in the \thdm\ case to 
\beq
C_{i}^2+C_{j}^2+C_{ij}^2=1\,,
\label{newsr}
\eeq
where $i\neq j$ are any two of the three possible indices.
The power of the Eq.~(\ref{newsr}) sum rule 
derives from the facts that it involves
only two of the neutral Higgs bosons and that
the experimental upper limit on any one $C_i^2$ derived from $\epem\to Z\h_i$
data is very strong: $C_i^2\leq 0.1$ for $m_{h_i}\leq 50$ GeV.
Thus, if $\h_i$ and $\h_j$ are both below about $50\gev$ in mass
then Eq.~(\ref{newsr}) requires that $C_{ij}^2\sim 1$,
whereas for such masses 
limits on $\epem\to \h_i\h_j$ from $\rts=161\gev$ and $172\gev$
data require $C_{ij}^2\ll 1$.
The excluded region in the $(m_{\h_i},m_{\h_j})$ plane that results
is illustrated in Fig.~\ref{fig95cl}
--- there cannot be two light
Higgs bosons even in the general CP-violating case.

The sum rule of Eq.~(\ref{newsr}) is required in order to have
unitary high energy behavior at tree-level
for the $\wp\wm\to\wp\wm$, $ZZ\to\wp\wm$ and $ZZ\to \h_i\h_i$
scattering amplitudes; see Eqs.~(4.1), (4.2) and (A.18)
of Ref.~\cite{wudka}, respectively.
In the context of a Higgs sector containing only 
doublet and singlet fields, the cited equations of Ref.~\cite{wudka}
reduce to the requirements 
\beq
 \sum_i C_i^2=1 \label{zzgensr} 
\eeq
\beq
 C_i^2+\sum_{k\neq i} C_{ik}^2=1\,,
\label{simpsr}
\eeq
where the 1 on the right hand side of Eq.~(\ref{simpsr}) arises from the 
$ZZ\h_i\h_i$ 4-point
interaction of Eq.~(\ref{zcoups}) contributing to $ZZ\to\h_i\h_i$ scattering. 
To derive Eq.~(\ref{newsr}) in the \thdm\ [for which $i,k=1,2,3$ 
in Eqs.~(\ref{zzgensr}) and (\ref{simpsr})],
we sum the $i=1,2$ and the $i=1,2,3$ cases of Eq.~(\ref{simpsr}),
respectively, to obtain
\bea
C_1^2+C_2^2+C_{12}^2&=&2-(C_{12}^2+C_{13}^2+C_{23}^2)\,,\label{zh2} \\
C_1^2+C_2^2+C_3^2&=&3-2(C_{12}^2+C_{13}^2+C_{23}^2)\,, \label{zh3}
\eea
respectively. We then employ the relation
$C_1^2+C_2^2+C_3^2=1$ from Eq.~(\ref{zzgensr})
in Eq.~(\ref{zh3}) to show that $C_{12}^2+C_{13}^2+C_{23}^2=1$
and substitute this result into Eq.~(\ref{zh2}) to obtain
$C_1^2+C_2^2+C_{12}^2=1$.  Cyclic permutation gives the other cases
of Eq.~(\ref{newsr}). 
Eq.~(\ref{newsr}) is much more useful for obtaining experimental
limits than either Eq.~(\ref{zzgensr}) or (\ref{simpsr})
since the latter two sum rules involve three Higgs
bosons (in the \thdm), whereas the former refers to just two.
This distinction only arises in the CP-violating case.
In the CP-conserving limit,
Eqs.~(\ref{zzgensr}) and (\ref{simpsr}) can be used to derive
Eqs.~(\ref{hHsr}) and (\ref{hAsr}), respectively, while Eq.~(\ref{newsr})
implies both (as described  earlier). 

These considerations can be generalized to extensions of the \thdm.
The simplest extension 
is to add one complex singlet (neutral) Higgs field. In this case,
it is no longer possible to place restrictions on two Higgs bosons.
The best that one can do is to 
write the sum rules in such a way as to demonstrate
that there cannot be three light neutral Higgs bosons.
Consider first the case where CP is conserved.  In the \tds\ model,
there will then be three CP-even Higgs bosons
(labelled 1,2,3 in order of increasing mass) and two CP-odd Higgs
bosons (labelled 4,5 in order of increasing mass). Using the
sum rules of Eqs.~(\ref{zzgensr}) and (\ref{simpsr}), with
$C_4=C_5=C_{12}=C_{13}=C_{23}=C_{45}=0$, we easily derive
the three crucial sum rules:
\bea
C_1^2+C_2^2+C_3^2&=&1 \,,\label{3even} \\
C_1^2+C_2^2+C_{14}^2+C_{24}^2&=&1+C_{35}^2 \,,\label{2even1odd}\\
C_1^2+C_{14}^2+C_{15}^2&=&1 \,.\label{1even2odd}
\eea
Eqs.~(\ref{3even}), (\ref{2even1odd}), and (\ref{1even2odd})
can be used to show that there cannot be three CP-even, two CP-even plus
one CP-odd,  and one CP-even plus two CP-odd Higgs bosons, respectively,
that are all light.
(In the second sum rule, since $C_{35}^2\geq 0$ the region within
which exclusion is certain is obtained by setting $C_{35}^2=0$.)
The boundaries in the three-dimensional mass spaces that follow
from these sum rules [with $C_{35}^2=0$ in Eq.~(\ref{2even1odd})] are shown
in Fig.~\ref{fig2d1scpc}. 

In the case that CP is violated,
the required sum rule for the \tds\ is obtained by generalizing the
procedure sketched for the derivation of Eq.~(\ref{newsr})
in the \thdm\ case.  Focusing on Higgs bosons numbers 1, 2 and 3,
one finds:
\beq
C_1^2+C_2^2+C_3^2+C_{12}^2+C_{13}^2+C_{23}^2=1+C_{45}^2\,.
\label{tdssr}
\eeq
This sum rule implies a lower bound (obtained with $C_{45}^2=0$)
for the sum of all the couplings squared responsible for production of
$ Z \h_1$, $ Z \h_2$, $ Z \h_3$, 
$\h_1\h_2$, $\h_1\h_3$ and $\h_2\h_3$ in $\epem$ collisions.
The portion of the $(m_1,m_2,m_3)$
mass space excluded by LEP data in the CP-violating case, as implied by
the sum rule of Eq.~(\ref{tdssr}), is shown in Fig.~\ref{fig2d1scpv}.

The above considerations can be further generalized to a Higgs sector
that contains $\ell$ doublets and $m$ neutral complex singlets; the number 
of physical neutral Higgs mass eigenstates is $2(\ell+m)-1$. The general sum
rule will apply to any subset containing $n=(\ell+m)$ of these Higgs
bosons. Let us label the members of the subset with indices $i=1,\ldots,n$.
Following the techniques illustrated in the \tds\ case, and assuming
that CP violation is present, we derive the coupling constant sum rule
\beq
\sum_{i=1}^nC_i^2+\sum_{\stackrel{\scriptstyle i,j=1}{i<j}}^n
C_{ij}^2=1+\sum_{\stackrel{ \scriptstyle i,j=n+1}{i<j}}^{2n-1}C_{ij}^2\,,
\label{gencase}
\eeq
where the most conservative bounds on the subset would be obtained
by setting all the $C_{ij}^2=0$ on the right hand side.
If CP violation is not present in the Higgs sector, then the above
sum rule will reduce to a simpler form that depends upon the CP
nature of the Higgs bosons included in the subset.

In this Letter, we have derived a new coupling constant sum rule
which makes it possible to use LEP data to exclude a
portion of $(m_{\h_1},m_{\h_2})$ mass space for the lightest
two neutral Higgs bosons of the most general
CP-violating two-Higgs-doublet model.
Although this region is not as large
as that excluded in the $(\mhl,\mha)$ mass parameter space
in the CP-conserving case, it is still very substantial.
Thus, LEP data implies that it is not possible for two of the three
neutral Higgs bosons of a general two-Higgs-doublet model to be light.
We have further shown how to extend this type of analysis to both
CP-conserving and CP-violating Higgs sectors
with an arbitrary number of doublets and singlets.
In the two-doublet plus one-singlet Higgs model,
LEP data already excludes 
the possibility that three of the five neutral Higgs bosons are light.

\vspace{2cm}
\centerline{\bf Acknowledgments}
\vspace{.5cm}
This work was supported in part by U.S. Department of Energy under
grants DE-FG03-91ER40674 and DE-FG03-92ER40689,
by the Davis Institute for High Energy Physics, by the
Committee for Scientific Research (Poland) under grant No. 2~P03B~180~09, 
and by Maria Sklodowska-Curie
Joint Fund II (Poland-USA) under grant No. MEN/NSF-96-252.
BG and JK would like to thank the Davis Institute for High Energy Physics for
hospitality.

\vskip .5in
\clearpage
{\it Appendix.} The limits presented in Figs.~\ref{fig95cl}-\ref{fig2d1scpv}
have been obtained from LEP1 and LEP2 data
on $\epem\to Z\h_i$ and $\epem\to \h_i\h_j$ production
using the following procedures. Consider first any given $Z\h_i$ channel.
For $m_{\h_i}<50\gev$, the 95\% CL upper limit on 
$C_i^2$ [defined in Eq.~(\ref{zcoups})]
is obtained by using the smaller of the values shown
in Fig.~5b of Ref.~\protect\cite{l3new} and Fig.~29 from
Ref.~\protect\cite{sopczak}. 
For $m_{\h_i}\geq 50\gev$, the 95\% CL upper limit on $C_i^2$ is
obtained as the ratio of the 95\% CL upper limit on the 
number of events as observed by ALEPH, taken to be
$\sim 3$ events from the graph in Ref.~\protect\cite{alephnew}
(which includes data at $\protect\rts=\mz$, $\protect\rts=161\gev$
and $\protect\rts\sim 172\gev$), to the number of events expected
at the given Higgs mass in the SM, as plotted in the same graph.
If the assumed $C_i^2$ exceeds the 95\% CL as defined above
at the input $m_{\h_i}$, then the parameter choice is taken to be excluded
at the 95\% CL. For a two-Higgs channel, $\h_i\h_j$, we approximate
the ALEPH 95\% CL limits implicit in Ref.~\protect\cite{alephnew}
by employing integrated luminosities of $L=11.08\pbi$ at $\protect\rts=161\gev$ 
and $L=10.5\pbi$ at $\protect\rts=172\gev$ and
the quoted efficiencies and branching ratios of $\eps=0.55,\br=0.83$
for the $4b$ channel
and $\eps=0.45,\br=0.16$ for the $2b2\tau$ channel. We compute the
expected number of events (combining the $4b$ and $2b2\tau$ channels)
for the input value of the $Z\h_i\h_j$ coupling strength
and then evaluate the Poisson probability that no events are observed.
If this probability is below 5\% then the chosen parameter set
for the $\h_i$ and $\h_j$ is said to be excluded at 95\% CL.

\begin{figure}[ht]
\leavevmode
\epsfxsize=2.8in
\centerline{\epsffile{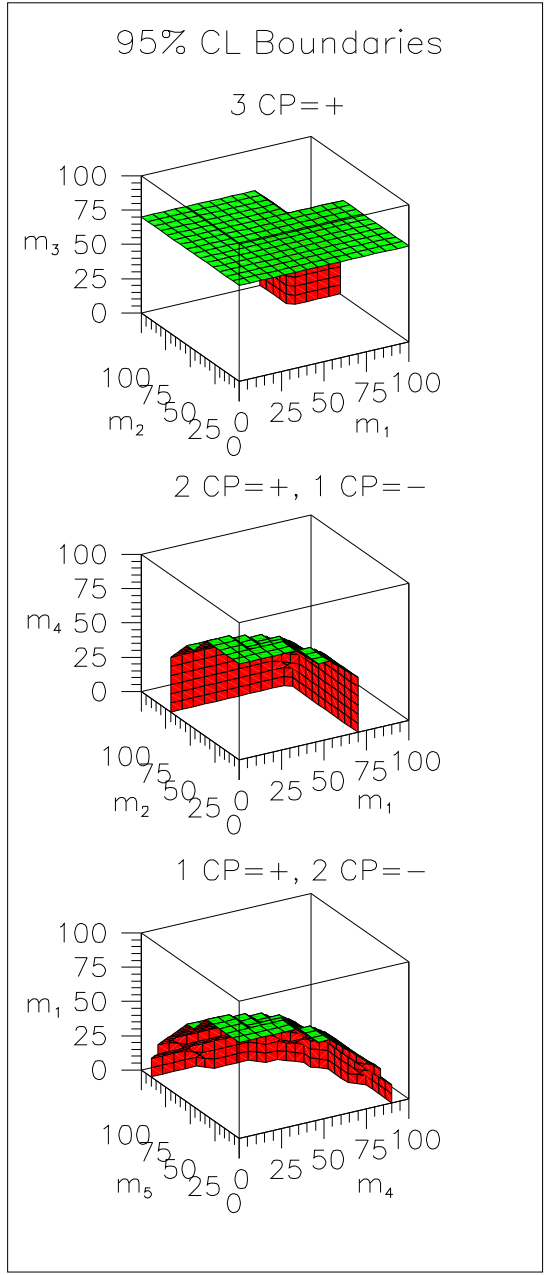}}
\bigskip
\caption{The LEP 95\% confidence level boundaries 
for the two-doublet plus one-singlet CP-conserving
Higgs sector in the $(m_1,m_2,m_3)$,
$(m_1,m_2,m_4)$ and $(m_4,m_5,m_1)$ spaces for
three CP-even, two CP-even plus one CP-odd, and two CP-odd plus one CP-even 
Higgs bosons, respectively. Mass axes are in GeV units.
Constraints are as in Fig.~\ref{fig95cl}.}
\label{fig2d1scpc}
\end{figure}

\begin{figure}[ht]
\leavevmode
\epsfxsize=3.0in
\centerline{\epsffile{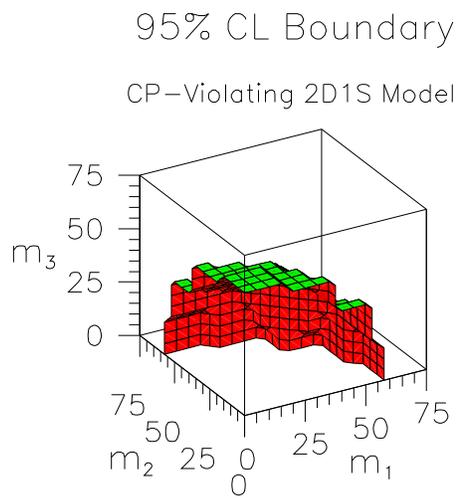}}
\bigskip
\caption{The LEP 95\% confidence level boundary
for the two-doublet plus one-singlet CP-violating
Higgs sector in the $(m_1,m_2,m_3)$ parameter space.
Mass axes are in GeV units. Constraints are as in Fig.~\ref{fig95cl}.}
\label{fig2d1scpv}
\end{figure}

\end{document}